\documentclass[letter]{aa}     

\usepackage{natbib}

\usepackage{graphicx}
\usepackage{txfonts}
\usepackage{lipsum}
\usepackage{subcaption}         
\usepackage{lscape}             
\usepackage{placeins}           

\newcommand{\planet}{LTT-9779 b}
\def\kms{$\mathrm{km\,s}^{-1}$}
\def\ms{\hbox{\,m\,s$^{-1}$}}         

\begin{document}

   \title{High-resolution detection of reflected light from the exo-Neptune \planet}

   \author{F.~Borsa\inst{1}
        }

   \institute{INAF -- Osservatorio Astronomico di Brera, Via E. Bianchi 46, 23807 Merate (LC), Italy}

   \date{Received 3 May 2026 / Accepted 18 June 2026}

  \abstract
{While high-resolution spectroscopy is routinely used to probe exoplanetary atmospheres, detecting reflected starlight remains highly challenging due to the extremely low planet-to-star optical flux ratios. We report the detection of reflected light from the ultra-hot exo-Neptune \planet\ using high-resolution spectroscopy with ESPRESSO in its 4UT mode. 
By combining multiple epochs and applying a cross-correlation analysis with a theoretical reflection kernel, we were able to identify a faint signal matching the expected morphological and kinematic profile of the planetary reflection. This signature, whose presence has been detected at a signal-to-noise ratio of $\sim$5.4, exhibits a radial velocity semi-amplitude consistent with the expected orbital motion.
We measured the planet-to-star flux ratio from the ratio of the equivalent widths of the planetary and stellar cross-correlation functions, finding $F_{\mathrm{p}}/F_\star = 102^{+29}_{-30}$ ppm for the ESPRESSO 380--770 nm wavelength range. Assuming a Lambertian phase function, this corresponds to a geometric albedo of $A_{\mathrm{g}} = 0.88 \pm 0.25$. The inferred albedo is consistent with previous space-based photometric measurements, suggesting a highly reflective atmosphere potentially dominated by scattering processes or high-altitude clouds. While we show that reflectivity is enhanced towards blue wavelengths, a detailed spectroscopic characterization of the planetary atmosphere from the reflected-light signal remains out of reach. This result highlights the scientific potential of future high-resolution spectrographs on extremely large telescopes, paving the way for systematic reflected-light detections across a broader exoplanet population, including cooler and smaller planets.
}


   \maketitle
\nolinenumbers

\section{Introduction}

The detection and characterization of exoplanet atmospheres now stand as the central goals of modern astrophysics. While transmission and thermal emission spectroscopy have provided crucial insights into the composition and structure of close-in and highly irradiated planets, the observation of starlight reflected by an exoplanet remains particularly challenging due to the extreme planet-to-star flux contrast.
The study of exoplanets in reflected light provides direct constraints on their geometric albedo, scattering properties, and atmospheric composition, particularly at optical wavelengths. Over the last decade, space-based photometry has successfully detected optical phase curves for a number of close-in exoplanets, enabling robust measurements of dayside brightness and global albedo. Thanks to the high photometric stability achieved by space missions, phase-curve variations at the level of a few parts per million can be routinely measured, making photometry a well-established tool for characterizing reflected light from exoplanets.
In contrast, the detection of reflected starlight using high-resolution spectroscopy (HRS) remains significantly more challenging. Although HRS offers the advantage of resolving individual spectral lines and exploiting the large Doppler shift of the planet to disentangle the planetary signal from stellar and telluric contributions \citep[see, e.g.,][]{Birkby2018,Snellen2025}, the extreme planet-to-star contrast in reflected light has limited the technique thus far. To date, HRS studies have primarily resulted in upper limits on the reflected-light signal or in tentative and debated detections \citep[e.g.,][]{CollierCameron1999, Charbonneau1999,Leigh2003,Rodler2010,Martins2015,Borra2018,Hoeijmakers2018,DiMarcantonio2019,Scandariato2021,Vaughan2026}, highlighting both the promise of the method and the difficulty of achieving a robust measurement.

Nevertheless, HRS holds unique potential for reflected-light studies. By coherently combining the signal from thousands of stellar absorption lines reflected by the planetary atmosphere, HRS has the capacity to directly constrain the planet’s wavelength-dependent albedo. This chromatic information is essential for probing atmospheric scattering properties, such as the presence of high-altitude clouds or hazes, and tracing its overall composition. As instrumental stability, collecting power, and data analysis techniques continue to improve, HRS is poised to become a key tool for the detailed spectroscopic characterization of exoplanetary atmospheres in reflected light, complementing space-based photometry and unlocking the direct spectroscopic characterization of a wide range of exoplanets.

In this work, we present the high-resolution spectroscopic detection of the reflected-light signal from the exo-Neptune \planet. To achieve this aim, we analyzed archival data gathered with the ESPRESSO spectrograph operating in its 4UT mode \citep{Pepe2021}, exploiting the combined collecting power of the Very Large Telescope to unveil the faint planetary signal.

\section{Observations and data reduction}

We downloaded from the ESO Science Archive Facility the raw ESPRESSO high-resolution spectrograph data relative to the 4UT dataset as analyzed in \citet{Vaughan2026}, corresponding to 230 spectra taken over three different observing nights, with an average signal-to-noise ratio (S/N) of $\sim$200 in order 60.
ESPRESSO covers a simultaneous wavelength range of 380-788 nm, with a spectral resolution of R$\sim$70,000 in its 4UT mode \citep{Pepe2021}.
The raw data were reduced with the dedicated pipeline (DRS v3.3.0\footnote{https://ftp.eso.org/pub/dfs/pipelines/espresso/espdr-pipeline-manual-3.3.10.pdf}) provided by ESO and the ESPRESSO Consortium \citep{Pepe2021}. 
In particular, we were interested in the DRS computation of the cross-correlation function (CCF) of the host star, which is standard in calculating precise stellar radial velocities. We calculated it using a binary mask of G8 stellar type for each observation.

The ultra-short orbital period of \planet\ ($\sim$0.79 days) induces a substantial radial velocity variation of $\sim$200 \kms\ across a single observing sequence. Furthermore, the apparent rotation of the host star, as seen from this planet, is expected to significantly broaden the reflected-light signal by $\sim$60 \kms\ \citep{Spring2022}. Thus, to try to recover a large exoplanetary signal which spans hundreds of \kms, we have to reduce the CCFs with a large width. Since the maximum width allowed in the DRS is $\pm$300 \kms, for each exposure, we reduced two different CCFs, each one with a step of 1 \kms. The first in the range [-511:89] \kms, and the second in the range [-111:489] \kms. The two CCFs were then stuck together, after verifying their full consistency on the overlapping velocities. In this way, we have for each observation a stellar CCF that spans the [-511:489] \kms\ range, centered on the $\sim$-11 \kms\ value of the systemic velocity, with step 1 \kms.

\section{Analysis and results}

\begin{figure}[h]
    \centering
    \includegraphics[width=8.5cm, trim={0cm 0.cm 0cm 1.2cm}, clip]{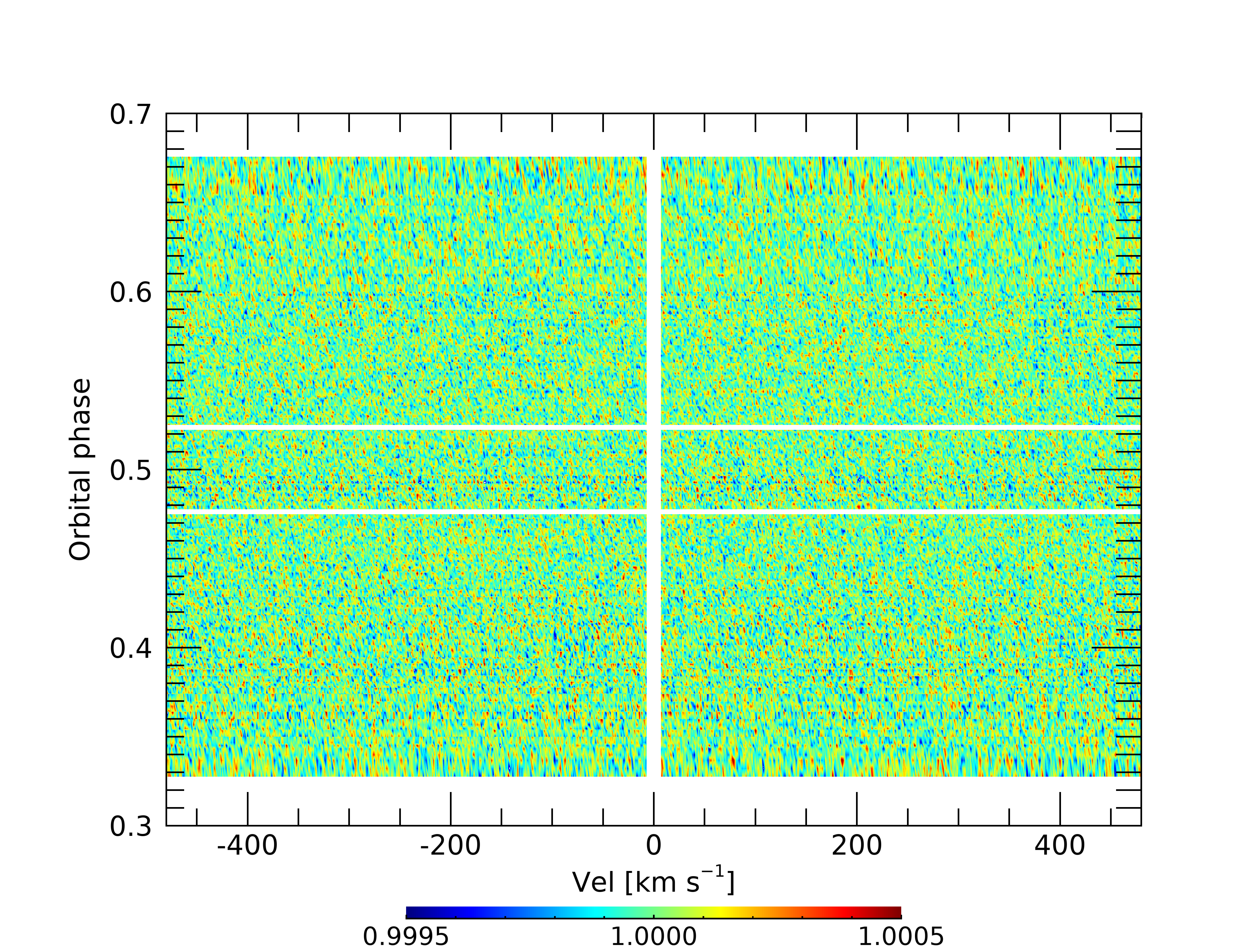}
    \caption{CCF residual map, after removing the stellar contribution. Horizontal white lines show the beginning and ending of the planetary eclipse. The masked region of the center of the stellar CCF residual can be noticed centered at Vel=0.}
    \label{fig:ccf_res}
\end{figure}

For each observation, we normalized the CCF and put it in the stellar restf rame, subtracting the theoretical stellar radial velocity (throughout this work we use the system parameters given in Table \ref{tabParameters}).
Then we calculated a master stellar CCF for each night separately by performing a weighted average of the nightly CCFs, similarly to what is usually done for emission spectroscopy \citep[e.g.,][]{Borsa2022}. The planetary signal does not affect the master, since the fast planetary movement ensures the change of its position during the observations. We then divided each single CCF by the master stellar CCF of the respective night and masked the region [-7:+7] \kms\ (the center of the stellar CCF), where the stellar CCF residual is stronger. The residual map CCF$_{res}$ (Fig.~\ref{fig:ccf_res}) has a standard deviation of $\sim$160 ppm, which is impressive and highlights the potential of ESPRESSO used in its 4UT mode. The planetary reflected-light signal is hidden in this map. We note that by tracing the reflected stellar CCF, our analysis does not extract the planetary atmospheric spectrum; rather, it measures the broadband geometric albedo of the planet averaged over the specific wavelength range.

\subsection{The expected planetary reflected-light signal}
\label{sect:model}
To search for the planetary reflected-light signal, we first examined its expected shape and amplitude to use it as a model to investigate the observational data. From previous photometric observations, we know that the overall planetary albedo is high. While optical measurements with CHEOPS ($\sim0.4-1.0~\mu\mathrm{m}$) delivered an albedo of $\sim$0.8 strictly from the secondary eclipse depth \citep{Hoyer2023}, here we adopted as a first guess the value of $A_{\mathrm{g}}$$\sim$0.5 measured by \citet{Coulombe2025} with JWST in the $\sim$$0.6-1.0~\mu\mathrm{m}$ range. Although our data were taken at shorter wavelengths, we preferred to take the latter as a reference because it is derived from the planetary phase curve, making it methodologically more comparable to our approach of integrating over the pre- and post-eclipse orbital phases. 
We note that the geometric albedo of close-in planets is often expected to increase toward shorter wavelengths due to Rayleigh scattering and the presence of high-altitude clouds or hazes, which efficiently scatter blue photons while molecular absorption suppresses reflectivity at longer wavelengths \citep[e.g.,][]{Marley1999,Evans2013}. Assuming reflected light, the expected planet-to-star flux ratio can be written as
\begin{equation}
\frac{F_{\mathrm p}}{F_\star} = A_g \left(\frac{R_p}{a}\right)^2 \Phi(\alpha),
\end{equation}
where $A_g$ is the geometric albedo, $R_p$ the planetary radius, $a$ the orbital separation, and $\Phi(\alpha)$ the phase function at phase angle $\alpha$. For a Lambertian sphere, we have
\begin{equation}
\Phi(\alpha) = \frac{\sin\alpha + (\pi - \alpha)\cos\alpha}{\pi}.
\end{equation}
The average orbital phase of our observational set, when excluding exposures taken during secondary eclipse and assuming perfect symmetry pre- and post-eclipse, is 0.411. We can then expect ${F_{\mathrm p}}/{F_\star}\sim 6 \times 10^{-5}$.
We also know that the planetary reflected-light signal should have a wide broadening of $\sim$60 \kms\ \citep{Vaughan2026}, because of the fast rotation of the star as seen from the planet \citep{Spring2022}.

Within the cross-correlation framework, the CCF acts as a linear proxy of the integrated line flux (when using the same mask). Therefore, the measured ratio between the equivalent widths (EW) of the planetary and stellar CCFs can be directly compared to the expected reflected-light contrast via
\begin{equation}
\frac{EW_{\mathrm p}}{EW_\star} \simeq \frac{F_{\mathrm p}}{F_\star}.
\label{eq:2}
\end{equation}
We fit the stellar master CCF with a Gaussian profile, obtaining EW=5.3705$\pm$0.0002. To establish a theoretical baseline for the planetary reflection signal in the residual CCF map, we generated a standard rotational broadening kernel assuming a linear limb-darkening (LD) law \citep{Gray2008}. For this template, we adopted the expected kinematic broadening of 60 \kms\ and a planetary EW of $3.22 \times 10^{-4}$, as derived from Equation~\ref{eq:2}. The linear LD coefficient was fixed to the stellar value of 0.71, computed across the ESPRESSO passband using the \texttt{LDTK} toolkit \citep{Parviainen2015}.

\subsection{Looking for reflected light}
\label{Sect:looking}

\begin{figure}[h]
    \centering
    \includegraphics[width=8.7cm,trim={0.cm 0.1cm 0.cm 1.0cm}, clip]{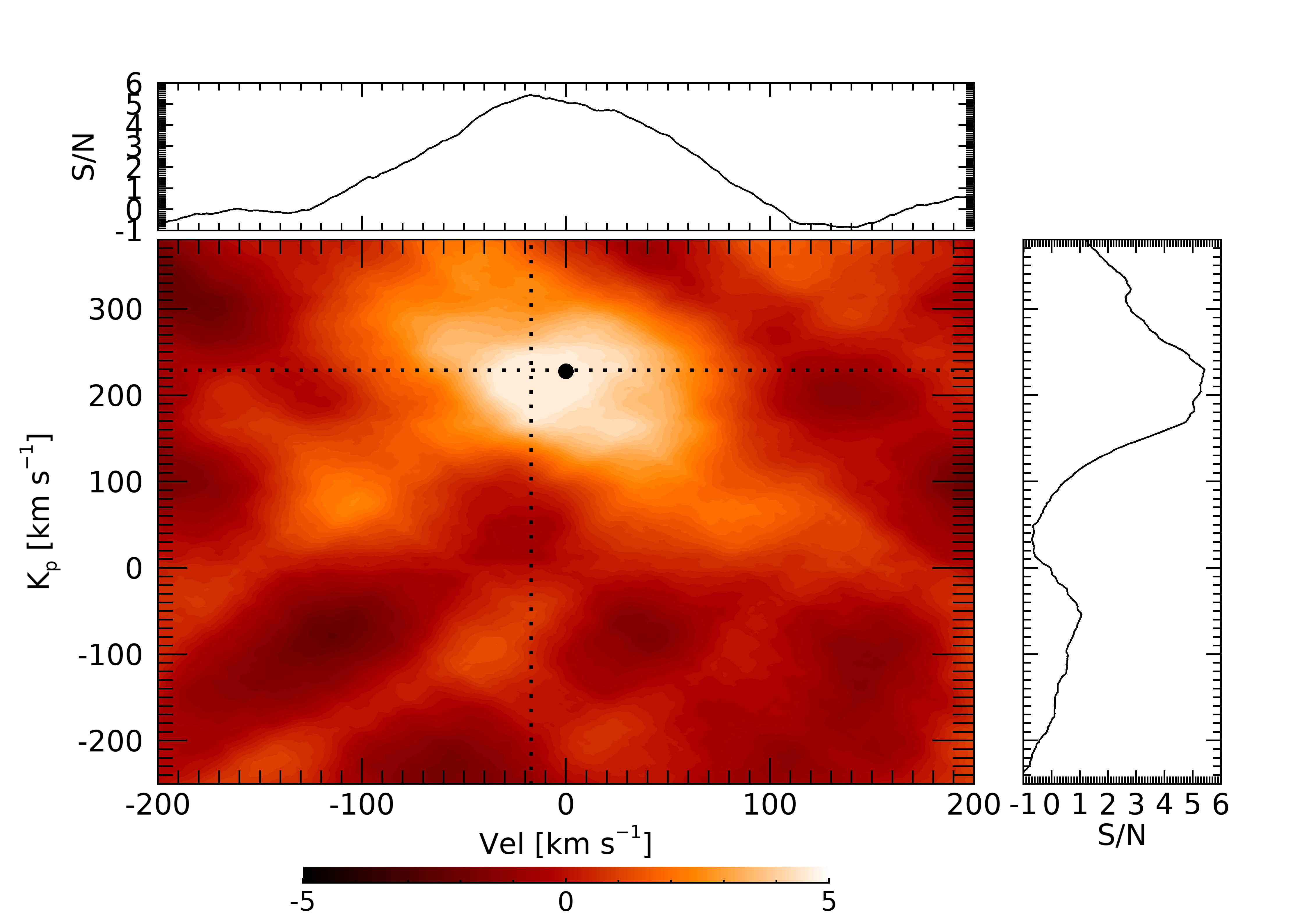}
    \caption{$K_{\mathrm{p}}$-$Vel$ map of the cross-correlation between the residual stellar CCFs and the expected planetary signal. The black dot marks the expected planet position. Dotted lines mark the peak of the map. The color scale shows the S/N of the detection.}
    \label{fig:kpvsys}
\end{figure}

We cross-correlated our model (Sect.~\ref{sect:model}) with each residual CCF$_{res}$, with step 1 \kms, and then pass to the $K_{\mathrm{p}}$-$Vel$ map, in the classic way it is routinely done for exoplanet atmospheric studies \citep[e.g.,][]{Snellen2010,Birkby2013}.
The CCFs were shifted to the planetary rest frame for a grid of trial values of the orbital velocity semi-amplitude ($K_{\mathrm{p}}$) and the shifted CCFs were summed in phase, enhancing any signal moving according to the planetary Keplerian velocity curve. The resulting 2D map is expected to peak at the planetary orbital parameters, if the expected signal is present.
We normalized the $K_{\mathrm{p}}$–$Vel$ map by the standard deviation of the off-signal regions, excluding the area around the expected planetary velocity and the map borders.
This provided an approximate S/N of the detection as a function of ($K_{\mathrm{p}}$, $Vel$).
We explored both positive and negative values of $K_{\mathrm{p}}$, to assess the robustness of the detection and to perform null tests against spurious correlations \cite[][]{Brogi2014}.
The results are shown in Fig.~\ref{fig:kpvsys}. There is a significant detection of the reflected-light signal with S/N=5.4, maximized at $K_{\mathrm{p}}$=$229^{+26}_{-64}$ \kms and $Vel$=$-17^{+46}_{-25}$ \kms. This position is perfectly compatible with the one expected for the planet ($K_{\mathrm{p}}$=$227\pm11$ \kms, $Vel$=0).

We note that the recovered error-bars on $K_{\mathrm{p}}$ and $Vel$ are quite large, which is caused by the fact that the planetary signal is weak and very broadened. To further validate the statistical significance of our detection, we performed additional checks, including a bootstrap analysis and an injection-retrieval test, described in detail in Appendix~\ref{sec:statistics}.

\subsection{Characterizing the reflected-light signal}
\label{Sect:fit}
Once we observed that the reflected-light signal is present and detected in the data, we tried to characterize it. We moved all the CCF$_{res}$ in the planetary restframe using the theoretical $K_{\mathrm{p}}$=$227.6$ \kms\ and averaged them, excluding all orbital phases in the range [0.4763:0.5237], representing when the planet is occulted by the star. 
We then performed a fit with a Gray rotational profile \citep{Gray2008}, employing a differential evolution Markov chain Monte Carlo (DE-MCMC) technique \citep{TerBraak2006, Eastman2013} in a Bayesian framework, running ten DE-MCMC chains of 50,000 steps and discarding the burn-in. 
In our initial exploratory fits, both the rotational broadening and the linear LD coefficient of the profile were left as free parameters. However, the LD remained unconstrained due to the low S/N, while the effective broadening was inflated to $\sim$90 \kms. 
 Since there is no obvious and plausible physical mechanism capable of justifying such an extreme broadening for this system, we investigated whether this overestimation could be driven by the low S/N using ad hoc simulations (Appendix~\ref{sec:statistics_broadening}). Our results confirm that this is the case, and the overestimation is likely due to profile-fitting spectroscopy under low S/N conditions, which skews the distribution toward broader solutions to minimize the $\chi^2$ penalty. 
To break this statistical degeneracy and extract a physically meaningful planetary signal, for our final analysis, we fixed the rotational broadening to the strictly kinematic expectation of 60 \kms. Furthermore, we fixed the LD coefficient to the stellar value of 0.71, avoiding the less realistic uniform disk approximation (a coefficient of 0). We left the Gray profile EW, center, and continuum as free parameters, setting uninformative priors. We only constrained the EW to be strictly positive (i.e., a positive albedo value), after verifying that relaxing this condition does not alter the statistical significance of our final results. The medians and the 15.86\% and 84.14\% quantiles of the posterior distributions were taken as the best values and $1\sigma$ uncertainties.

\begin{figure*}[h]
    \centering
    \includegraphics[width=17.4cm, trim={0.0cm 0.1cm 0.cm 0.4cm}, clip]{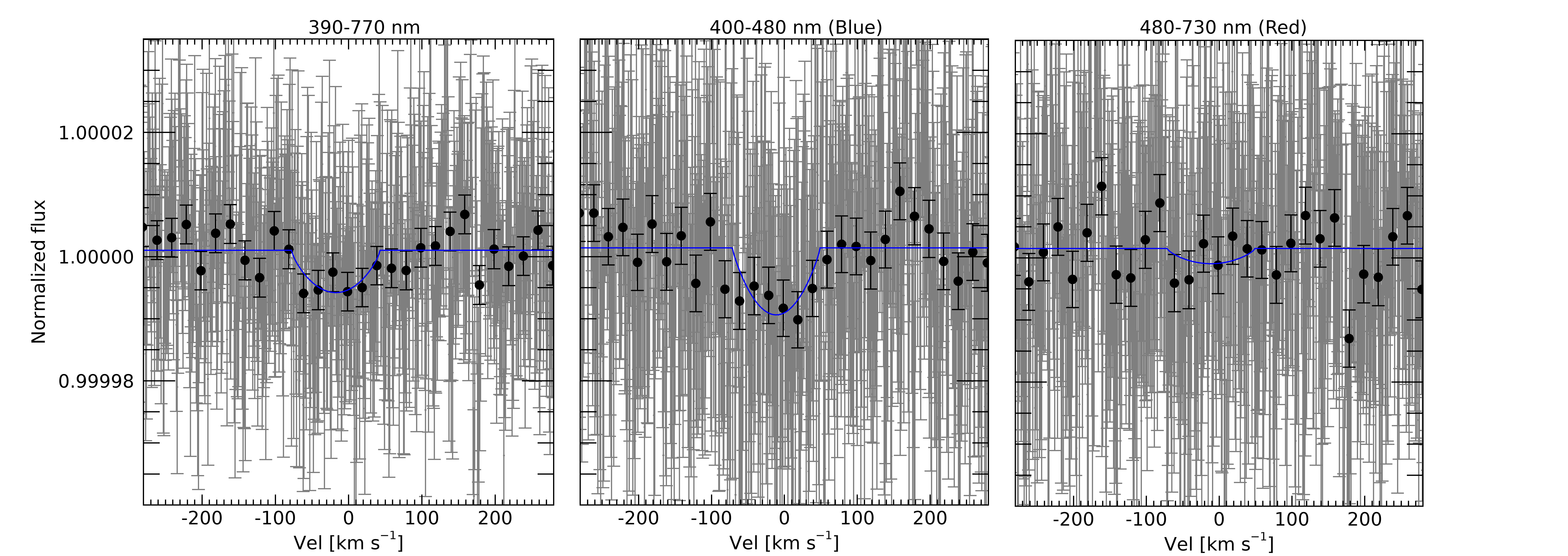}
    \caption{Reflected-light signal for the whole ESPRESSO wavelength range (left) as well as for the blue (middle) and red (right) bandpasses. Grey points show the original data, black points the 20 \kms\ binning. Blue line shows the best Gray profile fit.}
    \label{fig:reflected_ccf}
\end{figure*}

To determine the EW of the stellar CCF, we opted for direct numerical integration of the profile rather than relying on the Gaussian fit used in Sect.~\ref{sect:model} for our first-guess estimation.
This choice was made to minimize model-dependent approximations and to maintain methodological consistency throughout our study: in our subsequent chromatic analysis, the stellar CCFs in the individual spectral channels exhibited visible deviations from a purely Gaussian shape. 
The results are shown in Fig.~\ref{fig:reflected_ccf} and Table~\ref{tab:results}.
We detected a planetary signal with an EW=577$^{+165}_{-167}$ ppm, corresponding to ${F_{\mathrm p}}/{F_\star}$=102$^{+29}_{-30}$ ppm and a geometric albedo of $A_{\mathrm{g}}$=0.88$^{+0.25}_{-0.25}$, determined at S/N$\sim$3.5.

Given the successful detection of the reflected-light signal, we expanded our analysis to investigate potential spatial and spectral variations in the planetary atmosphere. Specifically, we performed two additional tests: first, to probe for longitudinal asymmetries, we divided our dataset into two subsets, analyzing the pre-eclipse and post-eclipse orbital phases independently.
Second, to assess the wavelength dependence of the reflected signal, we carried out a chromatic analysis. To this end, we split the ESPRESSO spectral range into two broad bins: a blue channel (echelle orders 10:35, $\sim$400-480 nm) and a red channel (orders 36:79, $\sim$480-730 nm). To ensure a comparable noise level across both channels, the split was chosen to yield similar dispersions in the resulting CCF residual maps ($\sim$238 and $\sim$236 ppm, respectively). Therefore, the division does not span equal wavelength ranges; rather, it compensates for the significantly higher density and depth of the stellar lines in the blue region of the ESPRESSO spectrum. The results are summarized in Fig.~\ref{fig:albedo} and Table~\ref{tab:results}.
The blue channel yields the only other detections (planetary albedo determined at S/N$\gtrsim$3.5), with a retrieved global albedo of $A_{\mathrm{g}} = 1.58^{+0.42}_{-0.42}$ (Fig.~\ref{fig:reflected_ccf}) and a post-eclipse albedo of $A_{\mathrm{g}} = 1.94^{+0.56}_{-0.57}$. We briefly discuss in Sect.~\ref{sec:chromatic} the signals extracted from the remaining spectral and orbital subsets.

\section{Discussion and conclusions}
\label{discussion}
In this work, we present the detection of reflected light from \planet\ using high-resolution spectroscopy. Our cross-correlation analysis successfully isolated the planetary signal, independently confirming previous indications of a highly reflective atmosphere \citep{Hoyer2023,Coulombe2025}. We observed a pronounced chromatic dependence, with the retrieved albedo rising steeply in the blue.
We note that a recent analysis of this same dataset by \citet{Vaughan2026} did not report a reflected-light detection. This apparent discrepancy can be explained by the different methodological choices of the two studies. To search for planetary atmospheric features, \citet{Vaughan2026} modeled the stellar spectrum as a continuum, a standard approach when dealing with expected large rotational broadening. However, by effectively removing or flattening the stellar absorption lines in their modeling, their analysis was inherently insensitive to the Doppler-shifted copy of the stellar spectrum. In contrast, our approach was specifically tailored to target and cross-correlate this reflected stellar signal, successfully recovering it. We further note that the use of the \texttt{sysrem} algorithm in their data reduction could have inadvertently suppressed a fraction of the planetary signal. Although the planetary trace shifts rapidly in radial velocity, its large rotational broadening makes it highly susceptible to partial subtraction by principal component analysis techniques.

Although the relative variations in our signal reliably trace the atmospheric properties of \planet, the absolute geometric albedos derived here are subject to methodological assumptions that can introduce systematic uncertainties. First, mapping the ratio of CCF EWs to a physical flux ratio ($F_{\mathrm{p}}/F_\star$) is complicated by differential line broadening and template contrast mismatches. Additionally, the CCF behavior may not scale perfectly linearly between the stellar and planetary regimes, which span vastly different orders of magnitude in terms of S/N. Second, the derivation of $A_{\mathrm{g}}$ assumes a uniform Lambertian phase function. However, previous observations \citep{Coulombe2025} and our hints for a morning/evening asymmetry explicitly break this assumption, meaning the true planetary reflectance could deviate from this simplified model. Finally, the observed chromatic gradient is predominantly driven by genuinely enhanced scattering efficiency in the blue; any geometrical effects of increased apparent planetary radius are likely negligible \citep[][]{Radica2024}.
It is worth noting the wavelength dependence of the retrieved planetary signature. While the detection in the full ESPRESSO passband is robust due to the lower noise floor, this integrated signal is dominated by the blue portion of the spectrum. When dividing the data chromatically, the reflection signature is recovered in the blue channel, whereas no significant signal is detected in the red channel, supporting a physical scenario where atmospheric reflectivity is sharply enhanced at shorter wavelengths.

Beyond the case of \planet\ presented here, this result represents a major step towards robust detection of reflected light at high spectral resolution and contributes to the ongoing effort to fully exploit this technique. It demonstrates that present-day HRS instruments, when operated at the limits of their performance, are already capable of detecting reflected light from low-mass, close-in exoplanets. This achievement lends strong confidence to the scientific potential of future high-resolution spectrographs on extremely large telescopes. In particular, ANDES on the ELT \citep{Marconi2024}, with its vastly increased collecting power and spectral coverage, will enable systematic reflected-light detections and detailed atmospheric studies of a much broader exoplanet population, including cooler and smaller planets \citep{Palle2025}.

\begin{acknowledgements}
We thank the referee for their important and constructive comments and suggestions that greatly improved the quality of the manuscript.
Based on data obtained from the ESO Science Archive Facility. We acknowledge the effort of the observing team who originally requested and executed these observations.
FB acknowledges support from Bando Ricerca Fondamentale INAF 2023 and INAF GO Large Grant 2024 "Strenghtening pylons for BRIDGES".
\end{acknowledgements}

\bibliographystyle{aa} 
\bibliography{bibliography.bib}

\begin{appendix}
\nolinenumbers
\section{Extended results}

\subsection{Looking for chromatic and longitudinal variations}
\label{sec:chromatic}
\begin{figure}[!h]
    \centering
    \includegraphics[width=9.5cm,trim={0.cm 0cm 0cm 0.cm}, clip]{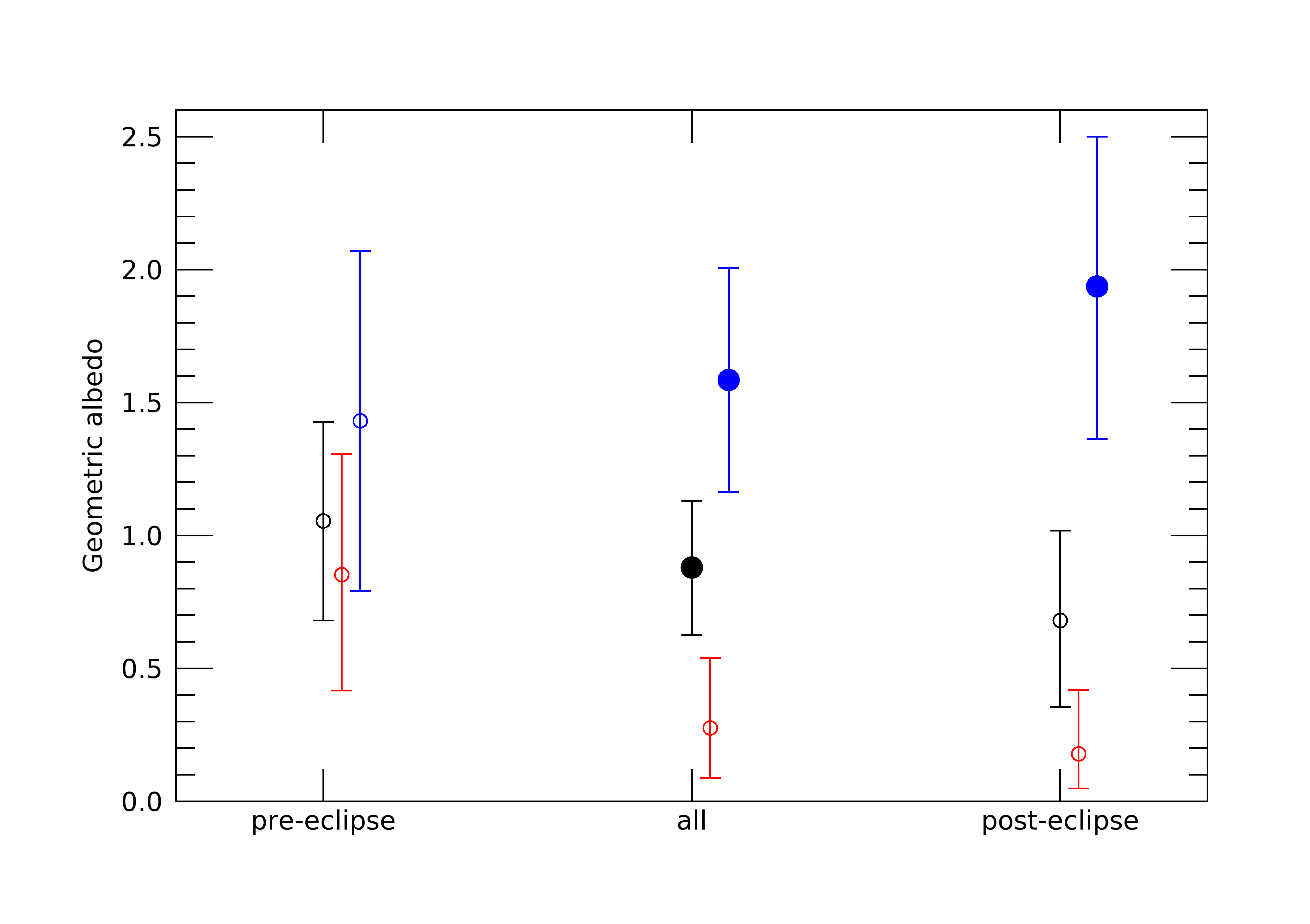}
    \caption{Measured planetary albedo at different orbital phases and different bandpasses. Black points refer to the overall ESPRESSO wavelength range ($\sim$390-770 nm), blue points to spectral orders 10:35 ($\sim$400-480 nm), red points to spectral orders 36:79 ($\sim$480-730 nm). Filled points show results with S/N$\gtrsim$3.5.}
    \label{fig:albedo}
\end{figure}

Our analysis performed in Sect.~\ref{Sect:fit} hints to a chromatic and longitudinal asymmetry across \planet's dayside, the latter of which was already evidenced by \citet{Coulombe2025}. Specifically, we see indications of a strong, blue-sloped reflected signal during the post-eclipse phases, corresponding to the highly reflective western dayside (Fig.~\ref{fig:albedo}). In contrast, the eastern dayside (probed pre-eclipse) yields a chromatically flat signal. This dichotomy may be attributed to localized cloud formation on the morning terminator. While the extreme heating of the dayside evaporates aerosols by the time atmospheric circulation reaches the eastern dayside (the evening terminator), the cooler nightside temperatures may allow thick clouds to condense and survive as they rotate onto the western dayside. These morning clouds dominate the optical reflection, producing the observed blue chromaticity.

\begin{figure}[!h]
    \centering
    \includegraphics[width=9.3cm, trim={0.0cm 0.cm 0.5cm 0.cm}, clip]{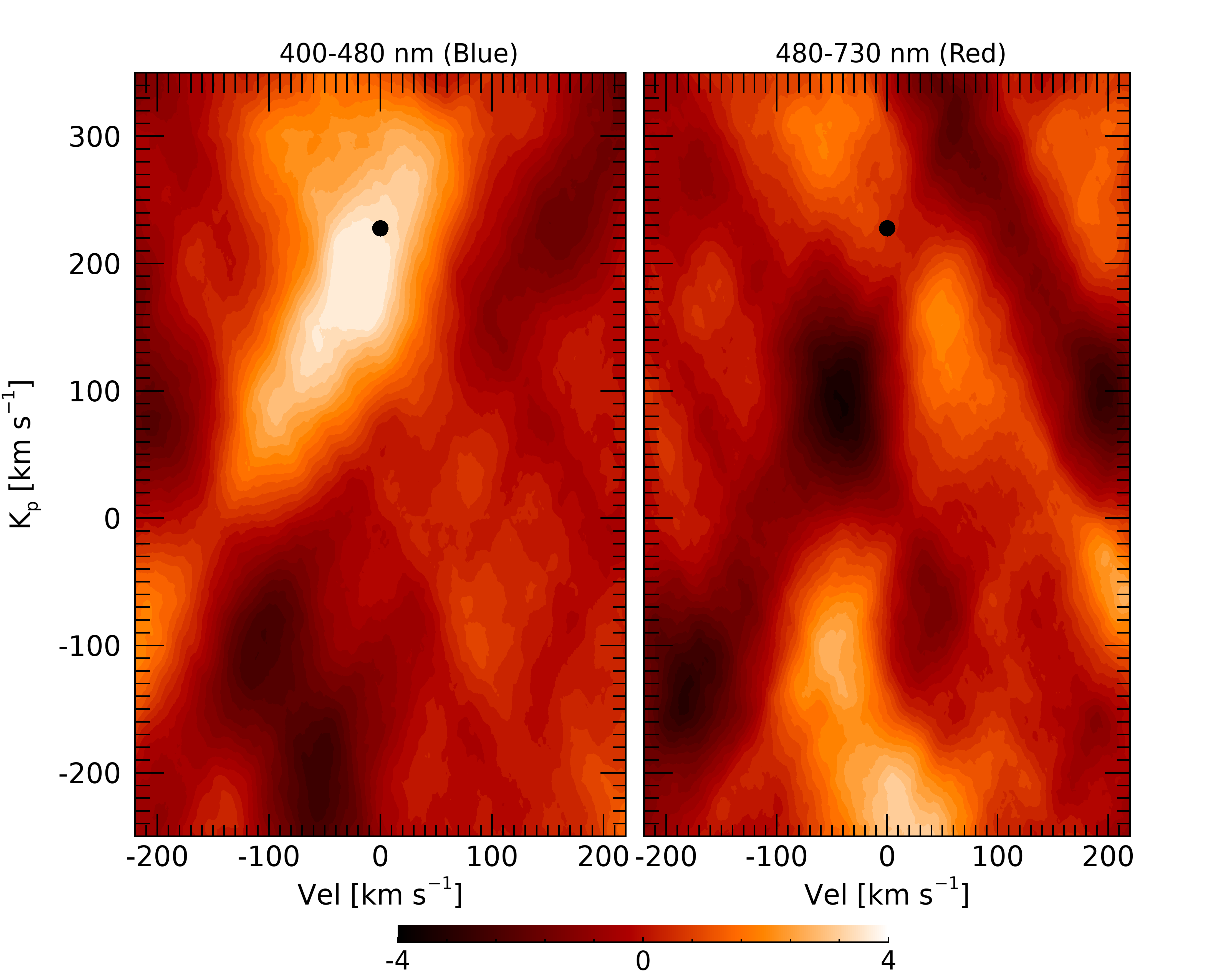}
    \caption{Chromatic $K_{\mathrm{p}}$-$Vel$ maps for the blue (left) and red (right) wavelength subsets. The black dots mark the expected planet position. Color scale shows the S/N of the detection.}
    \label{fig:kpvsys_chr}
\end{figure}

\begin{table}[!h]
\begin{center}
\caption{Properties of the LTT-9779 planetary system used in this work.}
\label{tabParameters}
\footnotesize
\begin{tabular}{ccc}
 \hline\hline
 \noalign{\smallskip}
 Parameter & Value & Reference\\
 \noalign{\smallskip}
 \hline
\noalign{\smallskip}
\multicolumn{3}{c}{LTT-9779}\\
\noalign{\smallskip}
\hline
\noalign{\smallskip}
\multicolumn{3}{c}{\it Stellar parameters}\\
\noalign{\smallskip}
\noalign{\smallskip}
$T_{\rm eff}$ [K] &   5480 $\pm$ 42 & \citet{Edwards2023}\\
$log g$ &  4.47 $\pm$ 0.11& \citet{Edwards2023}\\
$Fe/H$ & 0.25  $\pm$ 0.04& \citet{Edwards2023}\\
$M_{\rm s}$ [M$_{\sun}$] &1.02$\pm$0.02 & \citet{Edwards2023} \\
$R_{\rm s}$ [R$_{\sun}$] &0.949$\pm$0.006 &  \citet{Edwards2023}\\
$P_{\rm rot}$ [days] &45 &  \citet{Jenkins2020}\\
LD 380-770 nm & $0.71$&  This work\\
LD 400-480 nm & $0.86$&  This work\\
LD 480-730 nm & $0.67$&  This work\\
\noalign{\smallskip}
\noalign{\smallskip}
\multicolumn{3}{c}{\it Planetary and orbital parameters}\\
\noalign{\smallskip}
\noalign{\smallskip}
$Period$ [days]&  0.792064(10 $\pm$ 14) & \citet{Edwards2023}\\
$T_0$ [BJD] & 2459043.310(680$\pm$115) & \citet{Vaughan2026}\\  
$a$ [AU] &   0.01679 $\pm$ 0.00013 & \citet{Jenkins2020}\\
$e$ &   $0.0$ & \citet{Jenkins2020}\\
$i$ [degrees]&   76.4 $\pm$ 0.4 & \citet{Jenkins2020}\\
V$_{\rm sys}$ [\kms] & -11.0 & assumed \\
$M_{\rm p}$ [M$_{\rm Jup}$] & 0.09225 $\pm$ 0.0025 & \citet{Jenkins2020} \\
$R_{\rm p}$ [R$_{\rm Jup}$] & 0.421 $\pm$ 0.021 & \citet{Jenkins2020} \\
$K_{\rm s}$ [\ms] & 19.65 $\pm$ 0.43 & \citet{Jenkins2020}\\
\noalign{\smallskip}
 \hline
\end{tabular}
\end{center}
\end{table}

\begin{table*}[!h]
    \begin{center}
        \caption{Gray profile fitted parameters for the planetary signal and derived ${F_{\mathrm p}}/{F_\star}$ and geometric albedo.}
        \label{tab:results}
        \footnotesize
            \begin{tabular}{cccccc}
                \hline\hline
                \noalign{\smallskip}
                Range & Planet EW [ppm] & Shift [\kms] & Fixed broadening [\kms] & ${F_{\mathrm p}}/{F_\star}$ [ppm] & Albedo \\
                \noalign{\smallskip}                
                \hline
                \noalign{\smallskip}
All ($\sim$390-770 nm)&   577$^{+165}_{-167}$ & -17.5$^{+15.2}_{-9.1}$ &     60.0 &
   102$^{+29}_{-30}$ &      0.88$^{+ 0.25}_{-0.25}$\\
Pre-transit &   692$^{+ 244}_{- 246}$ &      -19.8$^{+ 20.9}_{-13.0}$ &       60.0 & 123$^{+ 43}_{- 44}$ &   1.05$^{+ 0.37}_{- 0.37}$\\
Post transit &   447$^{+ 222}_{- 215}$ &      -7.7$^{+ 26.9}_{- 18.9}$ &       60.0 &  79$^{+ 39}_{- 38}$ &    0.68$^{+ 0.34}_{- 0.33}$\\
Blue ($\sim$400-480 nm)&   890$^{+ 237}_{- 237}$ &      -11.2$^{+ 15.6}_{-  12.4}$ &   60.0 &  184$^{+  49}_{- 49}$ &       1.58$^{+ 0.42}_{- 0.42}$\\
Blue pre-eclipse &   803$^{+ 359}_{-  359}$ &      -30.5$^{+ 23.4}_{- 12.6}$ &       60.0 & 166$^{+ 74}_{- 74}$ &       1.43$^{+ 0.64}_{-0.64}$\\
Blue post-eclipse &    1087$^{+  317}_{- 322}$ & 2.5$^{+13.5}_{-14.2}$ &       60.0 & 225$^{+   65}_{- 67}$ &   1.94$^{+ 0.56}_{- 0.57}$\\
 Red ($\sim$480-730 nm)&   209$^{+  199}_{-   142}$ &      -11.2$^{+ 40.1}_{- 24.3}$ &       60.0 &
   32$^{+ 30}_{-  22}$ &   0.28$^{+  0.26}_{-  0.19}$\\
Red pre-eclipse &  647$^{+344}_{-  330}$ &      -3.7$^{+ 28.4}_{- 21.7}$ &       60.0&
   99$^{+ 52}_{- 51}$ &      0.85$^{+  0.45}_{-  0.44}$\\
Red post-eclipse &   135$^{+  182}_{-  98}$ &  -8.3$^{+ 43.1}_{- 30.3}$ &       60.0&
   21$^{+ 28}_{- 15}$ &     0.18$^{+  0.24}_{- 0.13}$\\
                \noalign{\smallskip}
                \hline
            \end{tabular}
    \end{center}
\end{table*}

\subsection{Chromatic $K_{\mathrm{p}}$-$Vel$ maps}
\label{sec:chromatic_kpvel}

To further investigate the wavelength dependence of the planetary signature, we computed the $K_{\mathrm{p}}$-$Vel$ maps as in Sect.~\ref{Sect:looking} separately for the two chromatic subsets. The planetary reflection signal is successfully recovered in the blue channel with S/N$\sim$4.5, whereas no significant peak is detected in the red channel (Fig.~\ref{fig:kpvsys_chr}). This contrast confirms that the detection in the full ESPRESSO passband is overwhelmingly driven by the blue spectral orders. While integrating over the entire wavelength range is statistically beneficial to lower the noise floor, the physical origin of the reflected light is clearly skewed toward shorter wavelengths.

\FloatBarrier

\section{Statistical tests}
\label{sec:statistics}

\subsection{Bootstrap}
\begin{figure}[h]
    \centering
    \includegraphics[width=\linewidth, trim={0cm 0.7cm 0.6cm 1.cm}, clip]{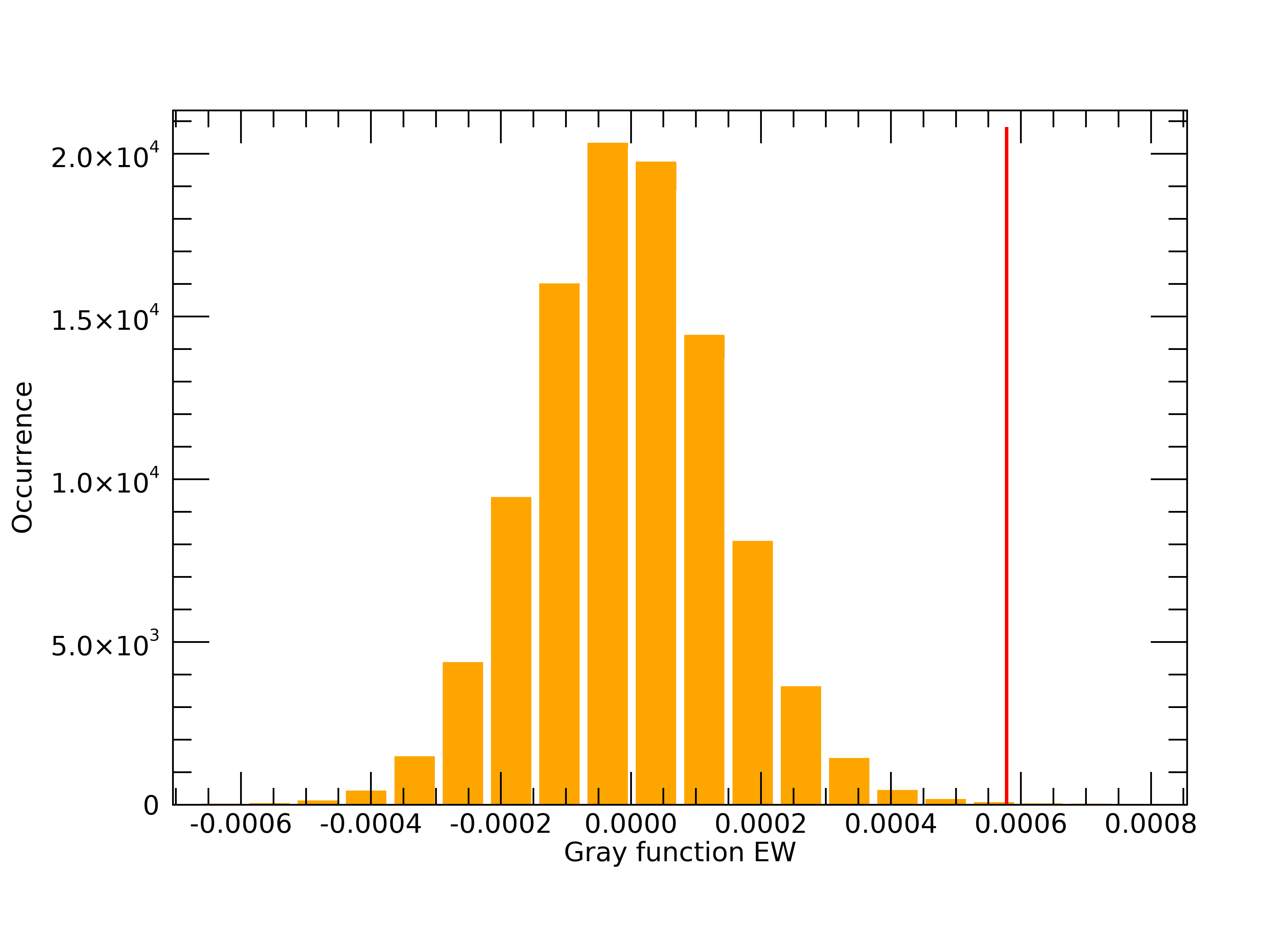}
    \caption{Random retrieved EWs distribution created with the bootstrap method. The vertical red line shows the EW of the reflected light detection.}
    \label{fig:bootstrap}
\end{figure}

To further assess the statistical significance of our white light reflected-light detection, we performed a bootstrap analysis, adapting the method proposed in \citet{Hoeijmakers2020} to our non-transiting dataset similarly as in \citet{Borsa2022}. We masked all regions where the planetary signal was expected (within $\pm40$ \kms\ from the theoretical planetary velocity), shifted every residual cross-correlation function CCF$_{res}$ in the time series by a random radial velocity drawn from a uniform distribution, and then averaged the resulting CCF$_{res}$. Next, we fitted Gray profiles with fixed stellar LD and broadening of 60 \kms\ centered at a random position within the averaged CCF. This procedure was repeated 100,000 times.
Figure \ref{fig:bootstrap} displays the resulting distributions of these random fluctuations, showing that the EW of our reflected light detection is significantly stronger. To estimate the significance of the detection, we fitted a Gaussian function to the random distribution to determine its standard deviation $\sigma$. We calculated the significance as the ratio between the EW of the reflected-light signal and this $\sigma$. The significance of the detection results at 4.1$\sigma$ (Fig.~\ref{fig:bootstrap}).

\subsection{Injection-retrieval}
\begin{figure}[h]
    \centering
    \includegraphics[width=\linewidth, trim={0.5cm 0.cm 0.3cm 1.cm}, clip]{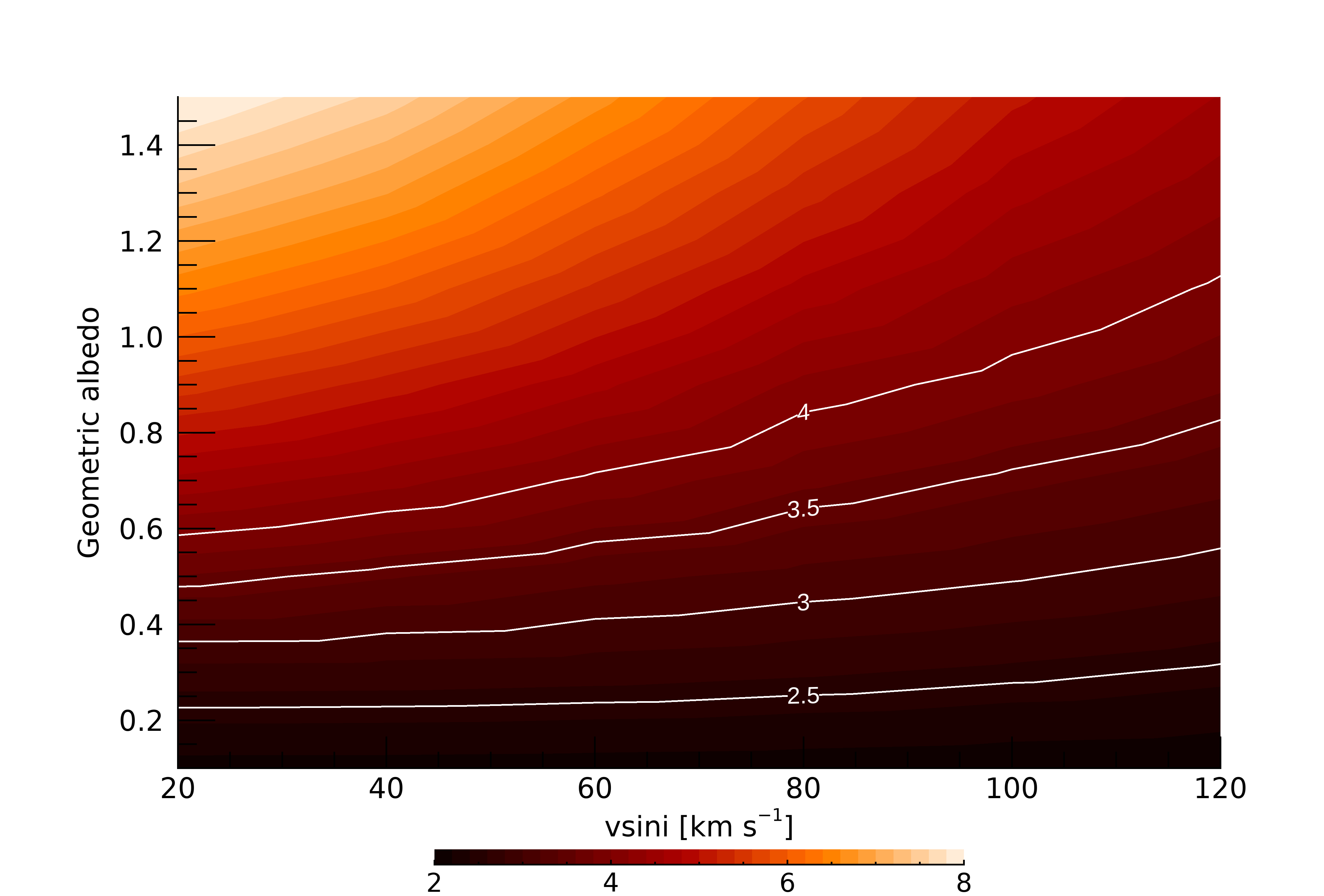}
    \caption{Injection-retrieval test of an expected reflected-light signal, performed on the global ESPRESSO dataset. White lines and color map show the S/N of the retrieved signal.}
    \label{fig:detection_limits}
\end{figure}

To evaluate our sensitivity and map the detection limits, we carried out an injection-retrieval test. A simulated reflected-light signal was injected directly into the pipeline-reduced CCFs, prior to any analysis. To avoid any bias or interference from the actual planetary signal present in the data, the mock signal was injected at the opposite Keplerian velocity semi-amplitude ($-K_{\mathrm{p}}$) and with an inverted sign. The synthetic signal was modeled assuming a Gray profile with linear LD (Sect.~\ref{sect:model}), scaling the albedo according to the planetary orbital phase (no signal injected during planetary eclipse). We performed this injection process across a grid of parameters, varying both the geometric albedo and the broadening. For each parameter combination, we extracted the detection significance of the retrieved signal from the corresponding $K_{\mathrm{p}}$–$Vel$ map (see Sect.~\ref{Sect:looking}). The resulting detection limits are presented in Fig.~\ref{fig:detection_limits}, which illustrates the retrieval significance as a function of the explored broadening and albedo values.

We remind that the approximation made in this work, also during this injection test, is that the ratio between the planetary and stellar CCF equivalent widths directly traces the reflected-light contrast ${F_{\mathrm p}}/{F_\star}$.
We note that the injected signal partially overlaps the one present in the data at orbital phases close to the eclipse. However, being of opposite sign, this may result in a suppression of the injected signal and not in enhancement, thus making this test conservative.
The measured planet-to-star flux ratio was converted into a geometric albedo $A_{\mathrm{g}}$ by assuming the planet behaves as a Lambertian sphere.

\subsection{Impact of low S/N on the broadening retrieval}
\label{sec:statistics_broadening}
\begin{figure}[h]
    \centering
    \includegraphics[width=\linewidth, trim={0.5cm 0.7cm 1cm 0.8cm}, clip]{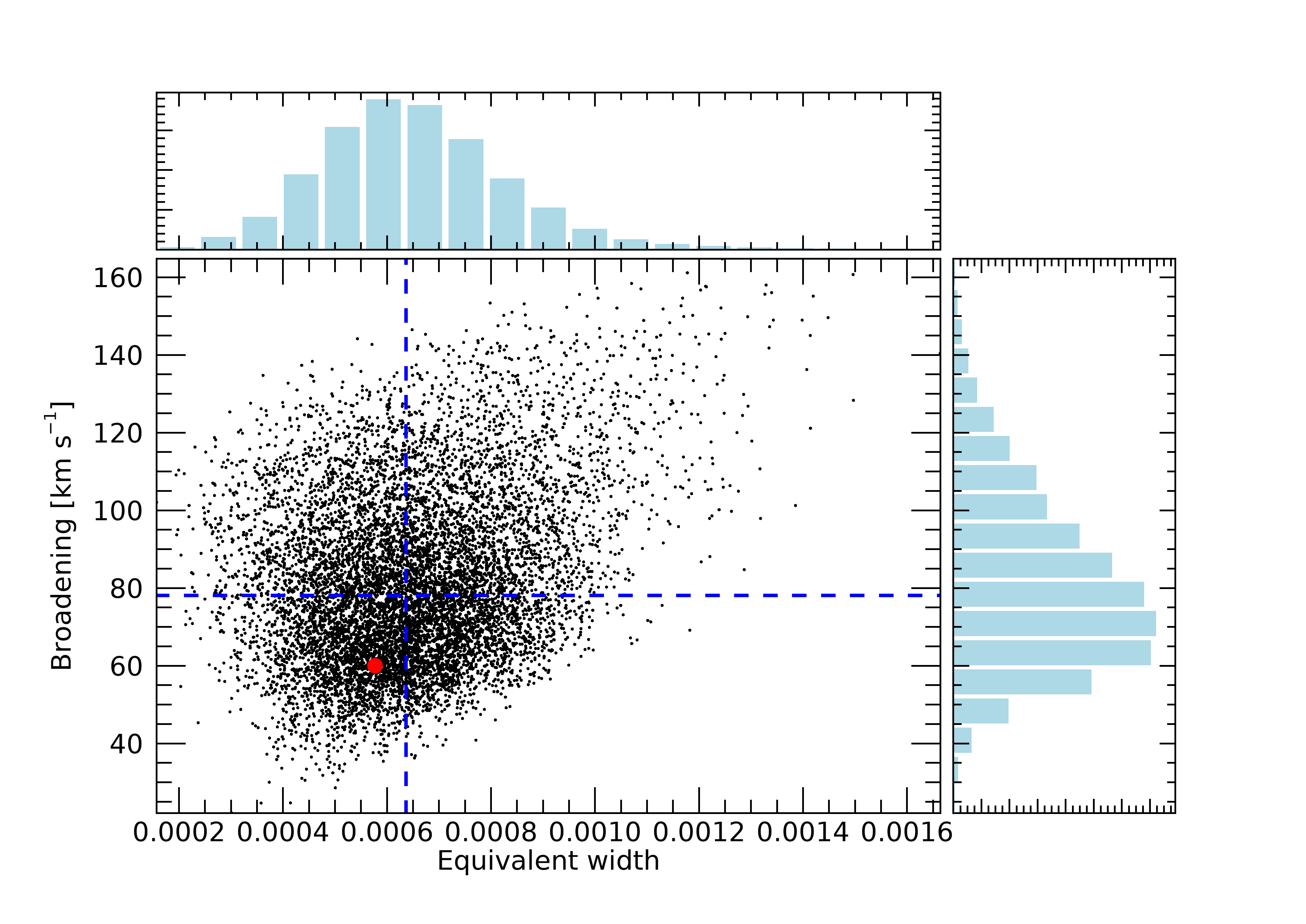}
    \caption{Retrieved broadening and EW values for the simulation performed. The red point shows the true values, while the blue dashed lines show the median of the distributions of the retrieved values. Both distributions result skewed towards large values, in particular the broadening one.}
    \label{fig:broadening_simu}
\end{figure}

To validate the hypothesis that the low S/N drives the inflated rotational broadening observed in the unconstrained fit (Sect. \ref{Sect:fit}), we performed an ad hoc simulation. 
We adopted the Gray profile best fit of Sect. \ref{Sect:fit}, with the fixed broadening of 60 \kms, and injected random Gaussian noise, scaling the noise amplitude to the standard deviation measured in our real residual CCF maps. The sampling and flux uncertainties were kept as in the real data.
We then fitted this dataset using the exact same DE-MCMC framework and parameter setup employed for the main analysis (Sect.~\ref{Sect:fit}), this time deliberately leaving the rotational broadening as a free parameter. This procedure, starting from random noise injection, was repeated 10,000 times.
The resulting posterior distribution of the retrieved best fit broadening (Fig.~\ref{fig:broadening_simu}) reveals a clear systematic bias, and the same happens for the broadening distributions of each DE-MCMC run. Rather than being symmetrically centered around the injected value of 60 \kms, the posterior distribution is heavily asymmetric and skewed toward larger values, exhibiting a pronounced tail at high velocities. The median of the retrieved broadening systematically overestimates the injected value. This behavior visually and mathematically confirms our hypothesis: in this low-S/N regime, random noise fluctuations in the profile wings allow the sampler to artificially inflate the width to encompass the noise. 
This simulation reproduces the behavior observed in our actual data and validates our choice to freeze the rotational broadening to its expected kinematic value to extract a physically meaningful planetary albedo, which would otherwise be systematically inflated.

\end{appendix}

\end{document}